\documentclass[a4paper,12pt]{article}
\usepackage{epsfig}
\date{\today}
\newcommand{\bmat}{\left(\begin{array}}
\newcommand{\emat}{\end{array}\right)}
\newcommand{\be}{\begin{equation}}
\newcommand{\ee}{\end{equation}}
\newcommand{\ba}{\begin{eqnarray}}
\newcommand{\ea}{\end{eqnarray}}

\def\lsim{\raise0.3ex\hbox{$\;<$\kern-0.75em\raise-1.1ex\hbox{$\sim\;$}}}
\def\gsim{\raise0.3ex\hbox{$\;>$\kern-0.75em\raise-1.1ex\hbox{$\sim\;$}}}

\def\be{\beta}

\def\nn{\nonumber}

\def\lsim{\raise0.3ex\hbox{$\;<$\kern-0.75em\raise-1.1ex\hbox{$\sim\;$}}}
\def\gsim{\raise0.3ex\hbox{$\;>$\kern-0.75em\raise-1.1ex\hbox{$\sim\;$}}}

\begin{document}

\renewcommand{\thefootnote}{\fnsymbol{footnote}}

\pagestyle{empty}
\rightline{SUSX-TH/01-008}
\vskip 1cm
\begin{center}
{\bf \large{ On the EDM Cancellations in D-brane models\\[10mm]}}
{S. Abel, S. Khalil, and O. Lebedev \\[6mm]}
\small{Centre for Theoretical Physics, University of Sussex, Brighton BN1
9QJ,~~U.~K.\\[7mm]}
\end{center}

\hrule
\vskip 0.3cm
\begin{minipage}[h]{14.0cm}
\begin{center}
\small{\bf Abstract}\\[3mm]
\end{center}

We analyze the possibility of simultaneous electron, neutron, and mercury 
 electric dipole moment (EDM) cancellations in the mSUGRA and D--brane models.  
We find that the mercury EDM constraint practically rules out the
cancellation scenario in D-brane models whereas in the context of mSUGRA
it is still allowed with some fine-tuning.

\end{minipage}
\vskip 0.2cm
\hrule
\vskip 1cm


One of the most important tests of CP-violation comes from
the measurements of the electric dipole moments (EDMs).
Non-observation of the EDMs imposes severe constraints on
models for physics beyond the Standard Model. 
  The most stringent of these come from continued
efforts to measure the EDMs
 of the neutron~\cite{bound},  electron~\cite{eedm},
and mercury atom \cite{mercury}
\begin{eqnarray} 
d_n &<& 6.3 \times 10^{-26} ~\mathrm{e~cm}~(90\% CL), \nonumber\\
d_e &<& 4.3 \times 10^{-27} ~\mathrm{e~cm}~,\nonumber\\
d_{Hg} &<& 2.1 \times 10^{-28} ~\mathrm{e~cm}~.
\end{eqnarray}
In particular, these constraints are  a difficult hurdle for 
supersymmetric theories if they are to allow sufficient baryogenesis. 
Indeed it is remarkable that the SM contribution to the EDM of the
neutron is of order $10^{-30}$ e cm, whereas the ``generic'' 
supersymmetric value is $10^{-22}$e cm. 

There are several proposals 
to reconcile the EDM constraints and supersymmetry. The EDM 
bounds may imply that the supersymmetric CP-phases are small \cite{smallphases}
 or  the sfermions of the first two generations are heavy \cite{heavy}.
Alternatively, they may imply that CP violation
has a flavor-off-diagonal character \cite{flavor}. It has also been
realized that in certain regions of the parameter space the constraints
on the CP-phases are not severe due to the EDM cancellations \cite{cancel}.
This last possibility will be the subject of our present study. 
In particular, it has recently been found that simultaneous  neutron
and electron EDM cancellations may occur in certain D-brane-motivated models \cite{lisa}
(see also \cite{brane}). 
We critically examine  theoretical aspects of the model of Ref.\cite{lisa} and 
analyze whether this
scenario  satisfies all of the experimental EDM constraints.

In our analysis, we follow the approach of Ibrahim and Nath \cite{cancel}, and 
 include contributions of the electromagnetic, chromomagnetic, and Weinberg
operators to the neutron EDM via Naive Dimensional Analysis
(a better justified approach to the NEDM based on the QCD sum rules
has recently appeared in \cite{pospelov1}). 
We have also included the Barr-Zee type contributions to the EDMs \cite{chang} and
the gluino-bottom-sbottom contribution to the Weinberg operator.
In addition 
to the electron and neutron EDM constraints, we impose the EDM constraint
for the mercury atom.
It has been realized that the mercury EDM is mostly sensitive to the
quark chromomagnetic dipole moments and that the constraint 
$d_{Hg} < 2.1 \times 10^{-28} ~\mathrm{e~cm}$ can be translated into \cite{pospelov}
\begin{equation}
\vert d_d^C - d_u^C -0.012 d_s^C \vert /g_s < 7 \times 10^{-27} cm\;,
\label{cedmlimits}
\end{equation}
where $g_s$ is the $SU(3)_c$ coupling constant and $ d_i^C$ are defined in the standard way
\cite{cancel}.
This constraint will be crucial in  our analysis. Before we proceed, let us briefly
review  basic ideas of the D-brane models (see also Refs.\cite{Ibanez:1999rf}
and \cite{carlos}).

Recent studies of type I strings have shown that it is possible to construct a number of 
models with non--universal soft SUSY breaking terms which are phenomenologically 
interesting. Type I models can contain $9$-branes, $5_i$-branes, $7_i$-branes, and 
$3$-branes where the index $i=1,2,3$ denotes the complex compact coordinate which is
included in the $5$-brane world volume or which  is orthogonal  to the $7$-brane world 
volume. However, to preserve $N=1$ supersymmetry in $D=4$ not all of these branes 
can  be present simultaneously and we can have (at most) either D9-branes with 
D$5_i$-branes or D3-branes with D$7_i$-branes.

Gauge symmetry groups are associated with stacks of branes located ``on top of each other''.
A stack of $N$ branes corresponds to the group $U(N)$. The matter fields are
associated with open strings which start and end on the branes. These strings may be attached
to either the same stack of branes or two different sets of branes which have
overlapping world volumes. The ends of the string carry quantum numbers associated
with the symmetry groups of the branes.
For example, the quark fields have to be  attached to the $U(3)$ set of branes,
while the  quark doublet  fields also have to be attached to the $U(2)$ set of branes.
Given a brane configuration, the Standard Model fields are constructed according to their
quantum numbers.

The SM gauge group can be obtained in the context of D-brane scenarios from 
$U(3)\times U(2)\times U(1)$, where the $U(3)$ arises from three coincident branes, 
$U(2)$ arises from two coincident D-branes and $U(1)$ from one D-brane. As explained
in detail in Ref.\cite{carlos},  there are different possibilities for 
embedding the SM gauge groups within these D-branes. It was shown that if the SM gauge 
groups come from the same set of D-branes, one cannot produce the correct values 
for  the gauge couplings $\alpha_j(M_Z)$ and the presence of additional matter (doublets
and triplets) is necessary to obtain the experimental values of the couplings 
\cite{Bailin:2000kd}. 
On the other hand, the assumption  that the SM 
gauge groups originate from different sets of $D$-branes leads in a natural 
way to intermediate values for the string scale $M_S \simeq 10^{10-12}$ GeV
\cite{carlos}. 
In this case, the analysis of the soft terms has been done under the 
assumption that only the dilaton and moduli fields contribute to supersymmetry breaking
and it has been found  that these soft terms  are generically non--universal. 
The MSSM fields arising from  open strings   are shown in Fig.1.
For example, the up quark singlets $u^c$ are states of the type $C^{9 5_3}$,
the quark doublets are $C^{9 5_1}$, etc.
The presence of extra ($D_q$) branes which are not associated with the SM gauge groups
is often necessary to reproduce the correct hypercharge and to cancel non-vanishing
tadpoles.

Recently there has been a considerable interest in supersymmetric models 
derived from D-branes \cite{lisa},\cite{brane}. 
In the model of Ref.\cite{lisa},  the gauge group
$SU(3)_c \times U(1)_Y$ was associated with  $5_1$ branes and  $SU(2)_L$ was associated with
$5_2$ branes. It was shown that in this model the gaugino masses are non--universal
($M_1=M_3 \ne M_2$) so that the physical CP phases 
 are $\phi_1 = \phi_3$, $\phi_A$ and $\phi_{\mu}$.  It was
emphasized that the non--universal gaugino phases have an important impact on enlarging
the regions of the parameter space where the EDM cancellations occur. 

However, closer inspection reveals that  such a model cannot produce the correct
hypercharge assignment for all of the SM fields. In fact, this model does not 
distinguish between the up and down quarks, $H_1$ and $H_2$, etc. which have 
different hypercharges and thus cannot be realistic. 
In realistic models, the hypercharge $U(1)$ is an anomaly-free  linear combination of two
or three $U(1)$'s arising from different sets of branes. As a result, 
the relation $\phi_1 = \phi_3$ can only be obtained if one embeds the SM gauge group
within the same set of branes. However, in this case the gaugino masses are universal
and the gaugino phase can be rotated away. The relation $\phi_1 = \phi_3 \not = \phi_2$,
which was found to be important  for the EDM cancellations, 
does not appear to  hold in realistic models.
Therefore in what follows, we will consider the EDM cancellations in both
the model of Ref.\cite{lisa} and a more realistic model. We shall find
that in both cases, although simultaneous EEDM and NEDM cancellations can occur
in considerable regions of the parameter space, imposing the mercury constraint
practically rules out the cancellation scenario.
  
Let us begin by constructing a more realistic D-brane scenario.
In order to obtain a model which is close in spirit to that of Ref.\cite{lisa},
one may place  the $U(1)$ brane on top of the $U(3)$ branes. This however implies
that the string scale $M_S$ is $6\times 10^8$ GeV which leads to 
$m_{3/2}\approx M_S^2/M_{Pl} \sim 10^{-1}$ GeV \cite{carlos},  too low a value for the 
SUSY particle masses. Extra light matter fields are required to mitigate this problem.
Another possibility is to consider a model without the $U(1)$ brane. This option
is also problematic since in this case the up-type Yukawa couplings are not allowed
\cite{carlos} resulting in a negligible top quark mass, whereas the lepton
Yukawa couplings are allowed. It is  difficult to imagine how this scenario can account
for  the observed fermion masses. Therefore, both of these simplified
versions of the model are hardly phenomenologically viable.

\begin{figure}[ht]
\begin{center}
\begin{tabular}{c}
\epsfig{file= 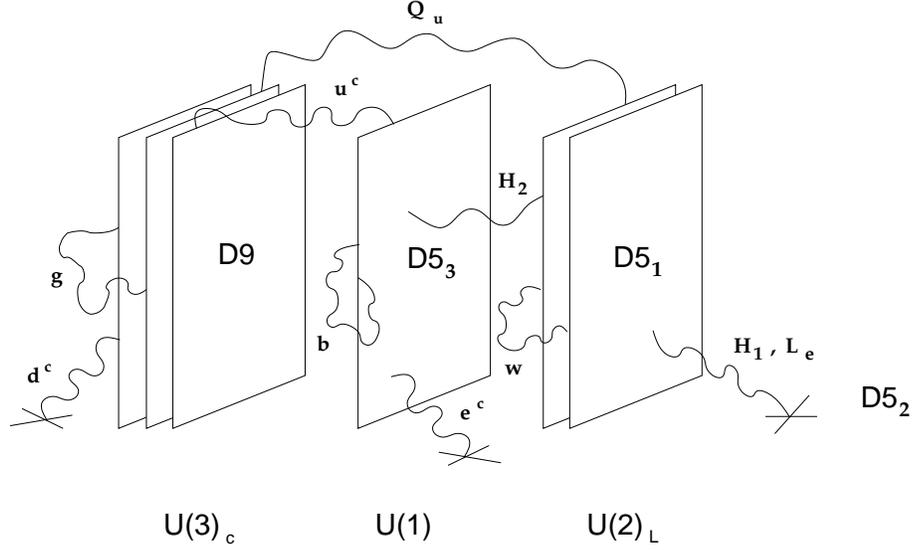, width=12cm}\\  
\end{tabular}
\end{center}
\caption{Embedding the SM gauge group within  different sets of D-branes. The extra
$D_q$ brane ($5_2$) is marked by a cross.}
\end{figure}

The model in which $U(3)$, $U(2)$, and $U(1)$ originate from different sets of branes
is much more phenomenologically attractive. In this case one naturally obtains an
intermediate string scale ($10^{10}-10^{12}$ GeV), although  higher values
up to $10^{16}$ GeV are still allowed. The Yukawa couplings are also more realistic:
both the up and the down type Yukawa interactions are allowed, while that for
the leptons typically vanishes (depending on further details of the model) \cite{carlos}.
The hypercharge is expressed in terms of the $U(1)$ charges $Q_{1,2,3}$ of the 
$U(1)_{1,2,3}$ groups:
\begin{equation}
Y=-{1\over 3} Q_3 -{1\over 2} Q_2 +Q_1\;,
\end{equation}
with the following $(Q_3,Q_2,Q_1)$ charge assignment:
\begin{eqnarray}
&& q=(1,-1,0) \;, \; u^c=(-1,0,-1) \;,\; d^c=(-1,0,0)\;,\\ \nonumber
&& l=(0,1,0) \;, \; e^c=(0,0,1) \;,\;\\ \nonumber
&& H_2=(0,1,1)\;,\; H_1=(0,1,0) \;.
\end{eqnarray}

Using the standard parameterization \cite{Ibanez:1999rf}:
\begin{eqnarray}
&& F^S= \sqrt{3} (S+S^*) m_{3/2} \sin \theta \;e^{-i\alpha_s} \;, \nonumber \\
&& F^i= \sqrt{3} (T_i+T^*_i) m_{3/2} \cos \theta\; \Theta_i e^{-i\alpha_i} \;,
\end{eqnarray}
and setting $\Theta_3=0$ for simplicity,
the gaugino masses in this model can be written as 
\begin{eqnarray}
M_3 & = & \sqrt{3} m_{3/2} \sin \theta \; e^{-i\alpha_s} \ ,  \\
M_{2} & = & \sqrt{3}  m_{3/2}\ \Theta_1 \cos \theta\; e^{-i\alpha_1} \ , \nn\\
M_{Y} & = &  \sqrt{3}  m_{3/2}\ \alpha_Y (M_S) 
\left( \frac{1}{\alpha_2 (M_S)}\Theta_1 \cos \theta e^{-i\alpha_1}
+\frac{2}{3\alpha_3 (M_S)}\sin \theta e^{-i\alpha_s}
\right) \;, \nonumber
\label{gaugino1}
\end{eqnarray}
where 
\begin{equation}
\frac{1}{\alpha_Y(M_S)} =
\frac{2}{\alpha_1(M_S)} + \frac{1}{\alpha_2(M_S)}
+ \frac{2}{3 \alpha_3(M_S)}\ .
\label{couplings}
\end{equation} 
Here $\alpha_k$ correspond to the gauge couplings of the $U(k)$ branes.
As  shown in Ref.\cite{carlos},  $\alpha_1(M_S) \simeq 0.1$ leads 
to the string scale $M_S \approx 10^{12}$ GeV.
Note that $\phi_3=\phi_Y$ if $\Theta_1=0$; this is however phenomenologically
unacceptable since in this case $M_2=0$ and the chargino is too light.
The soft scalar masses are given by
\begin{eqnarray}
m^2_{q} & = & m_{3/2}^2\left[1 -
\frac{3}{2}  \left(1 - \Theta_{1}^2 \right)
\cos^2 \theta \right] \ , \nn \\
m^2_{d^c} & = & m_{3/2}^2\left[1 -
\frac{3}{2}  \left(1 - \Theta_{2}^2 \right)
\cos^2 \theta \right] \ , \nn \\
m^2_{u^c} & = & m_{3/2}^2\left[1 -
\frac{3}{2} 
\cos^2 \theta \right] \ , \nn \\
m^2_{e^c} & = & m_{3/2}^2\left[1- \frac{3}{2}
\left(\sin^2\theta + \Theta_{1}^2 \cos^2\theta  \right)\right] \ , \nn \\
m^2_{l} & = & m_{3/2}^2\left[1- \frac{3}{2}
\sin^2\theta \right] \ , \nn \\
m^2_{H_2} & = & m_{3/2}^2\left[1- \frac{3}{2}
\left(\sin^2\theta + \Theta_{2}^2 \cos^2\theta  \right)\right] \ , \nn \\
m^2_{H_1} & = & m^2_l \;,
\label{scalars1}
\end{eqnarray}
and the trilinear parameters are
\begin{eqnarray}
A_{u} & = &  \frac{\sqrt 3}{2}m_{3/2}
   \left[\left(\Theta_{2} e^{-i\alpha_2} - \Theta_1 e^{-i\alpha_1} 
  \right) \cos\theta
- \sin\theta \;e^{-i\alpha_s} \right] \ ,
\label{trintrin}
\\
A_{d} & = &  \frac{\sqrt 3}{2}m_{3/2}
  \left[ -\left( \Theta_1 e^{-i\alpha_1} 
  +\Theta _{2} e^{-i\alpha_2}  \right) \cos\theta
- \sin\theta \;e^{-i\alpha_s} \right] \ ,
\label{trintrintrin}
\\
A_{e} & = &  0 .
\label{trilin11}
\end{eqnarray}
We note that the Yukawa couplings in Type I models are either 0 or 1, so an additional mechanism
is needed to produce the observed femion masses and mixings. 

In our EDM analysis, we rotate away the phase of $M_2$ by a $U(1)_R$ transformation
( the phase of $B\mu$ can also be set to zero by a $U(1)_{PQ}$ rotation).
We observe that the angles $\Theta_i$ and $\theta$ are quite constrained if we are to avoid
negative mass-squared's for squarks and sleptons.
For definiteness we assume $\alpha_1=\alpha_2$. 
Then the soft terms are parameterized in terms of the phase $\phi \equiv \alpha_1-\alpha_s$.

In Fig.2 we display the bands allowed by the electron (red), neutron (green), and
mercury (blue) EDMs. 
In this figure,
we set $m_{3/2}=150$ GeV, $\tan\beta=3$, $\Theta_1^2=\Theta_2^2=1/2$, 
$\cos^2\theta=2\sin^2\theta=2/3$,
and $\alpha_1(M_S) \sim 1$ with $M_S$ being the GUT scale.
For the plot to be more illustrative, we do not impose any additional constraints
besides the EDM ones (i.e. bounds on the chargino and  slepton masses, etc.).
It is clear that even though simultaneous EEDM/NEDM cancellations allow 
the phase $\phi$ to be  ${\cal{O}}(1)$, an addition of the  mercury constraint
 requires all phases to be very small (modulo $\pi$) and thus
practically rules out the cancellation scenario in this context.
 We find that the mercury
EDM behaviour in D-brane models  is very different from that of the electron and neutron
and thus is crucial in constraining the parameter space.

Next we consider the model of Ref.\cite{lisa}. As we have argued above, this
model can hardly be obtained from D-branes, however one may treat it as an interesting
phenomenological scenario motivated by D-branes. 
The (corrected) soft terms for this model read (for $\Theta_3=0$)
\begin{eqnarray}
&& M_Y=M_3=-A=\sqrt{3}m_{3/2} \cos\theta\; \Theta_1 e^{-i \alpha_1}\;, \nonumber\\
&& M_2=\sqrt{3} m_{3/2} \cos\theta\; \Theta_2 e^{-i \alpha_2}\;,\nonumber\\
&& m^2_L= m_{3/2}^2 (1-{3\over 2} \sin^2\theta);\,\nonumber\\
&& m^2_R= m_{3/2}^2 (1-3 \cos^2\theta ~\Theta_2^2)\;.
\end{eqnarray}
To illustrate the EDM constraints, 
we choose the parameters which allow for simultaneous EEDM/NEDM cancellations,
namely  $m_{3/2}=150$ GeV, $\tan\beta=2$, $\Theta_1=0.9$, 
and $\theta=0.4$ as given in Ref.\cite{lisa}. Fig.3 shows that the mercury constraint has the
same behaviour as in the model considered above and rules out large CP-phases.

Finally, we examine the possibility of simultaneous EDM cancellations in mSUGRA.
In contrast to the D-brane models, all of the EDM constraints have  similar 
behaviour in the $(\phi_A,\phi_{\mu})$ plane. In fact they can be approximately
described by the relation $\phi_{\mu} \simeq -a\sin\phi_A$ with $a>0$.
With a favorable choice of the parameters, the three bands will have a
significant overlap. In Fig.4 we present points allowed by 
all three EDM constraints with   all the masses
set to  200 GeV, $\tan\beta=3$, and $A=40$ GeV. Although $\phi_A$ is unconstrained
in this case, the phase $\phi_{\mu}$ is required to be ${\cal{O}}(10^{-2})$.
To relax this bound one has to either increase
the mass scale of the susy spectrum or restrict the range of $\phi_A$.
Our mSUGRA results are in agreement with those of Refs.\cite{pospelov} and \cite{Barger:2001nu}.

These results reveal that, putting aside the fine-tuning issues,
the EDM cancellation scenario is much more favored in the
mSUGRA framework than in the D-brane models. This qualitatively
agrees with the analysis of Ref.\cite{Barger:2001nu} where it was found
that in mSUGRA one  out of every $10^2$ points in the parameter space
satisfies the EDM constraints, while for the non-universal case
this fraction drastically reduces to $1/10^5$. 

To conclude, we have analyzed a possibility of simultaneous electron, neutron, and
mercury EDM cancellations in D-brane models and mSUGRA. We find that 
such cancellations cannot occur in presently available  semi-realistic D-brane models,
while in the mSUGRA framework these cancellations are allowed with some fine-tuning.

The authors are grateful to D. Bailin and C. Mu\~{n}oz for helpful communications,
and to M. Brhlik, M. Gomez, and T. Ibrahim for disscussion of the numerical
results.

\begin{figure}[ht]
\begin{center}
\begin{tabular}{c}
\epsfig{file= 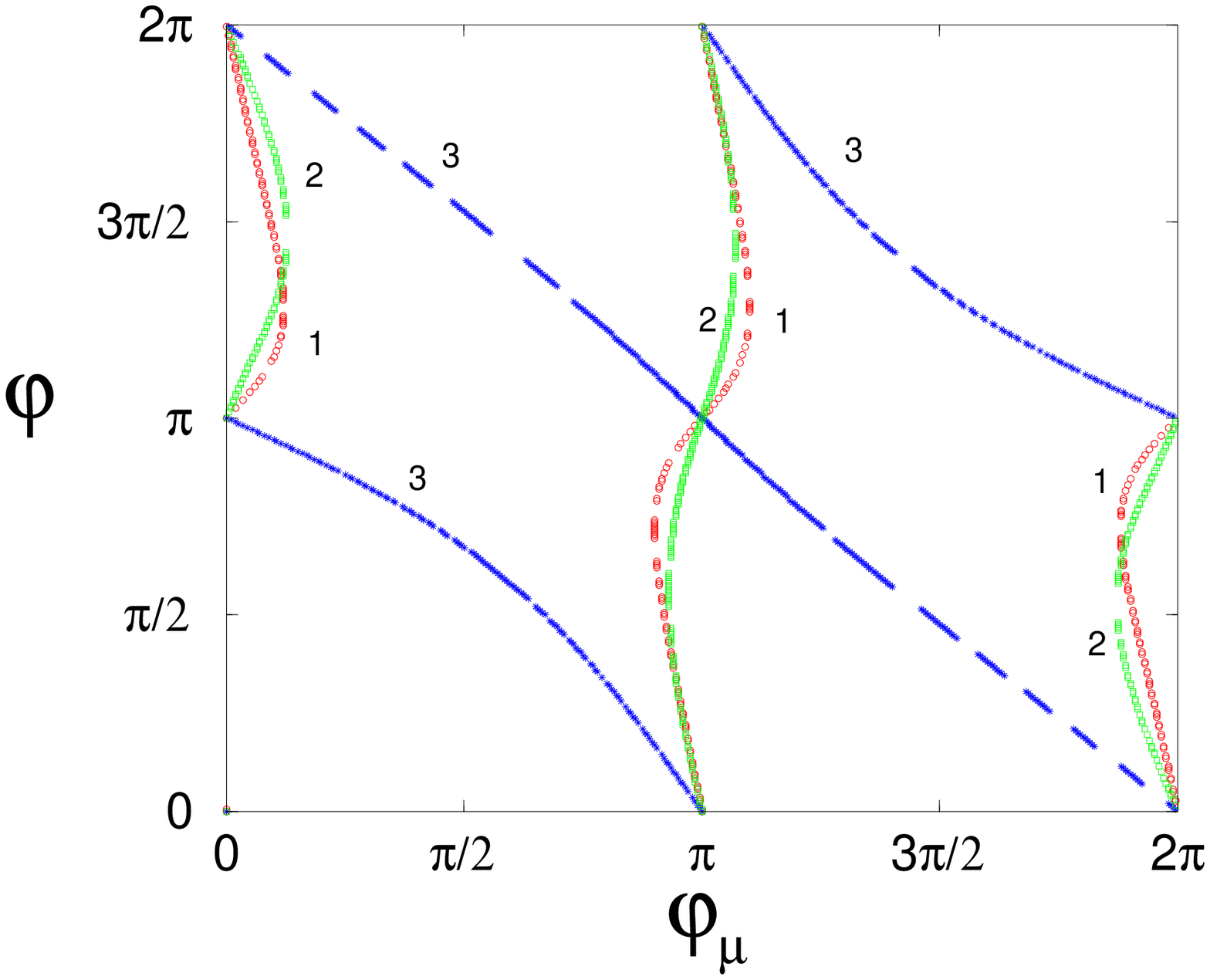, width=12cm}\\
\end{tabular}
\end{center}
\caption{ Bands allowed by the electron (1), neutron (2), and
mercury (3) EDMs in the D-brane model. The corresponding SUSY parameters
are given in the text.  }
\end{figure}
\begin{figure}[ht]
\begin{center}
\begin{tabular}{c}
\epsfig{file= 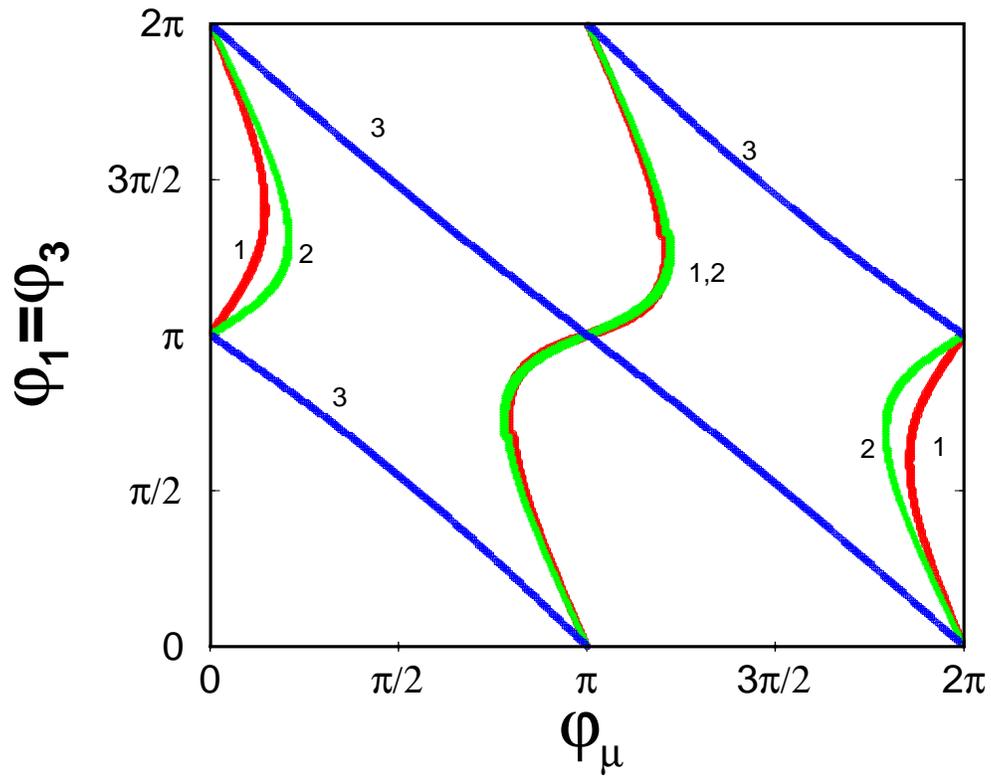, width=12cm}\\
\end{tabular}
\end{center}
\caption{Bands allowed by the electron (1), neutron (2), and
mercury (3) EDMs for the model of Ref.\cite{lisa}. The corresponding SUSY
parameters
are given in the text.  }
\end{figure}
\begin{figure}[ht]
\begin{center}
\begin{tabular}{c}
\epsfig{file= 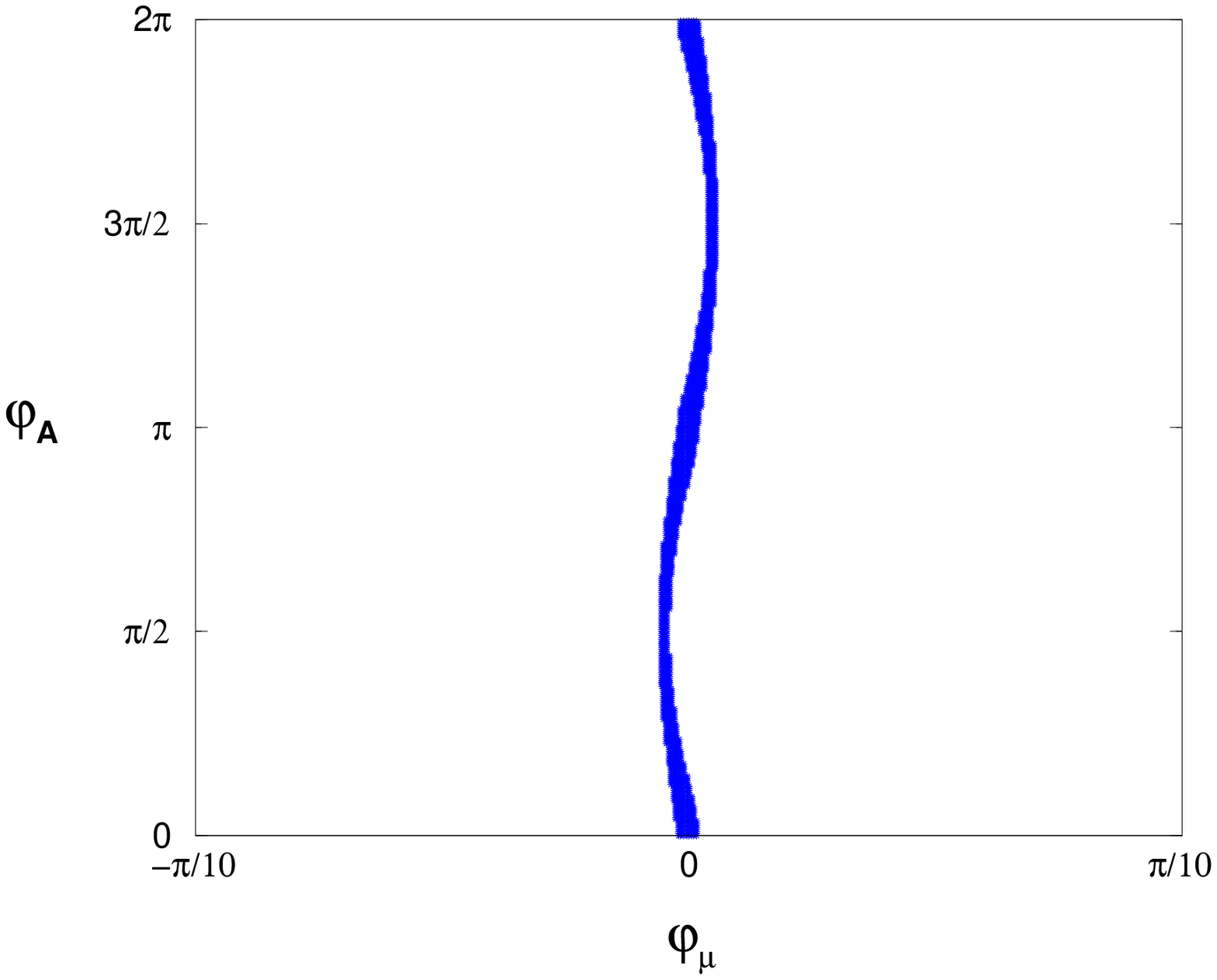, width=12cm}\\
\end{tabular}
\end{center}
\caption{Points allowed by simultaneous electron, neutron, and mercury
EDM cancellations in mSUGRA. The corresponding SUSY parameters are given in the text.}
\end{figure}


\begin{thebibliography}{99}

\bibitem{bound} P.G. Harris {\it et al.}, Phys. Rev. Lett. {\bf 82} (1999), 904;
        see also the discussion in S.K. Lamoreaux and R. Golub, Phys. Rev.  {\bf D61} (2000),
         051301.
\bibitem{eedm} E.D. Commins {\it et al.}, Phys. Rev. {\bf A50} (1994), 2960.
\bibitem{mercury} M.V. Romalis, W.C. Griffith, and E.N. Fortson, hep-ex/0012001;
         J.P. Jacobs {\it et al.}, Phys. Rev. Lett. {\bf 71} (1993), 3782.

\bibitem{smallphases} J. Ellis, S. Ferrara, D.V. Nanopoulos, Phys. Lett. B {\bf 114}, 
231 (1982); W. Buchm\"{u}ller and D. Wyler, Phys. Lett. B {\bf 121}, 321 (1983);
J. Polchinski and M.B. Wise, Phys. Lett. B {\bf 125}, 393 (1983).

\bibitem{heavy} P. Nath, Phys. Rev. Lett. {\bf 66} (1991), 2565; Y. Kizukuri and N. Oshimo,
        Phys. Rev. {\bf D46} (1992) 3025.

\bibitem{flavor} S.~Abel, D.~Bailin, S.~Khalil and O.~Lebedev, hep-ph/0012145, 
        to appear in Phys. Lett. {\bf B};
       R.~N.~Mohapatra and A.~Rasin, Phys.\ Rev.\ {\bf D 54}, 5835 (1996). 
       See also 
         S.A. Abel and J.-M. Fr\'{e}re, Phys. Rev. D {\bf 55}, 1623 (1997);
      S. Khalil, T. Kobayashi and A. Masiero, Phys. Rev. D {\bf 60}, 075003 (1999);
      S. Khalil and T. Kobayashi, Phys. Lett. B {\bf 460}, 341 (1999);
      S. Khalil, T. Kobayashi and O. Vives, Nucl. Phys. B {\bf 580}, 275 (2000).
       
\bibitem{cancel} T. Falk and K.A. Olive, Phys. Lett. B {\bf 375}, 196 (1996);
         Phys. Lett. B {\bf 439}, 71 (1998);
         T. Ibrahim and P. Nath, Phys. Rev. {\bf D57} (1998), 478;
        Errata-{\it ibid.} {\bf D58}, 019901 (1998); {\it ibid.} {\bf D60}, 019901 (1999);
        Phys. Lett. {\bf B418}, 98 (1998); Phys. Rev.  {\bf D58}, 111301 
        (1998); Erratum-{\it ibid.} {\bf D60}, 099902 (1999);
         M. Brhlik, G.J. Good and G.L. Kane, Phys. Rev. D {\bf 59}, 115004 (1999);
        A. Bartl, T. Gajdosik, W. Porod, P. Stockinger, and H. Stremnitzer,
        Phys. Rev. {\bf D60} (1999), 073003;
         S. Pokorski, J. Rosiek and C.A. Savoy, Nucl. Phys. B {\bf 570}, 81 (2000).
\bibitem{lisa} M. Brhlik, L. Everett, G.L. Kane and J. Lykken, Phys. Rev. Lett. {\bf 83}, 2124 (1999);
          Phys.\ Rev.\ D {\bf 62}, 035005 (2000).
\bibitem{brane}
           E.~Accomando, R.~Arnowitt and B.~Dutta, Phys.\ Rev.\ D {\bf 61}, 075010 (2000);
           T.~Ibrahim and P.~Nath, Phys.\ Rev.\ D {\bf 61}, 093004 (2000).

\bibitem{pospelov1} M. Pospelov and A. Ritz, hep-ph/0010037.

\bibitem{chang}
   D.~Chang, W.~Keung and A.~Pilaftsis,
   Phys.\ Rev.\ Lett.\ {\bf 82}, 900 (1999); Erratum-{\it ibid.} {\bf 83}, 3972 (1999).

\bibitem{pospelov} T. Falk, K.A. Olive, M. Pospelov, R. Roiban, Nucl.\ Phys.\ {\bf B560} (1999), 3.
\bibitem{Ibanez:1999rf} L.~E.~Ib\`{a}\~{n}ez, C.~Mu\~{n}oz and S.~Rigolin,
                 Nucl.\ Phys.\ B {\bf 553}, 43 (1999);
                 A. Brignole, L.~E.~Ib\`{a}\~{n}ez,  and C.~Mu\~{n}oz,
                 Nucl.\ Phys.\ B {\bf 422}, 125 (1994).
\bibitem{carlos} D.G. Cerde\~{n}o, E. Gabrielli, S. Khalil, C. Mu\~{n}oz, E. Torrente-Lujan,
          hep-ph/0102270.
\bibitem{Bailin:2000kd} G.~Aldazabal, L.~E.~Ib\`{a}\~{n}ez, F.~Quevedo and A.~M.~Uranga,
                    JHEP{\bf 0008}, 002 (2000);
                    D.~Bailin, G.~V.~Kraniotis and A.~Love, hep-th/0011289. 

\bibitem{Barger:2001nu}
V.~Barger, T.~Falk, T.~Han, J.~Jiang, T.~Li and T.~Plehn,
hep-ph/0101106.

\end{thebibliography}
\end{document}